\documentclass[final,3p,times]{elsarticle}

\usepackage{amsthm}
\usepackage{amsmath}
\DeclareMathOperator*{\argmax}{arg\,max}
\usepackage{listings}
\usepackage{float}
\usepackage{enumitem}
\usepackage{tcolorbox}
\usepackage{caption}
\usepackage{booktabs}
\usepackage{multirow}
\usepackage{array}
\newdefinition{definition}{Definition}
\newcolumntype{L}[1]{>{\raggedright\let\newline\\\arraybackslash\hspace{0pt}}m{#1}}
\newcolumntype{C}[1]{>{\centering\let\newline\\\arraybackslash\hspace{0pt}}m{#1}}
\newcolumntype{R}[1]{>{\raggedleft\let\newline\\\arraybackslash\hspace{0pt}}m{#1}}

\journal{Journal of Systems and Software}

\begin{document}

\begin{frontmatter}



\title{On the Impact of Multiple Source Code Representations on Software Engineering Tasks - An Empirical Study}

\author[iitt]{Karthik Chandra Swarna\corref{cor1}}
\ead{cs17b026@iittp.ac.in}
\cortext[cor1]{Corresponding Author}
\author[iitt]{Noble Saji Mathews}
\ead{ch19b023@iittp.ac.in}
\author[iitt]{Dheeraj Vagavolu}
\ead{cs17b028@iittp.ac.in}
\author[iitt]{Sridhar Chimalakonda}
\ead{ch@iittp.ac.in}
\affiliation[iitt]{organization={Research in Intelligent Software \& Human Analytics (RISHA) Lab, Department of Computer Science and Engineering, \\ Indian Institute of Technology Tirupati},
            country={India}}

\begin{abstract}
Efficiently representing source code is crucial for various software engineering tasks such as code classification and clone detection. Existing approaches primarily use \textit{Abstract Syntax Tree} (AST), and only a few focus on semantic graphs such as \textit{Control Flow Graph} (CFG) and \textit{Program Dependency Graph} (PDG), which contain information about source code that AST does not. Even though some works tried to utilize multiple representations, they do not provide any insights about the costs and benefits of using multiple representations. The primary goal of this paper is to discuss the implications of utilizing multiple code representations, specifically AST, CFG, and PDG. We modify an AST path-based approach to accept multiple representations as input to an attention-based model. We do this to measure the impact of additional representations (such as CFG and PDG) over AST. We evaluate our approach on three tasks: \textit{Method Naming}, \textit{Program Classification}, and \textit{Clone Detection}. Our approach increases the performance on these tasks by \textbf{11\%} (F1), \textbf{15.7\%} (Accuracy), and \textbf{9.3\%} (F1), respectively, over the baseline. In addition to the effect on performance, we discuss timing overheads incurred with multiple representations. We envision this work providing researchers with a lens to evaluate combinations of code representations for various tasks.
\end{abstract}


\end{frontmatter}


\section{Introduction}
Due to the drastic rise in the availability of enormous volumes of source code in open-source projects and tools to extract and analyze them \cite{bajracharya2014sourcerer, spadini2018pydriller, reza2020modelmine}, researchers have explored many ways of solving software engineering problems such as \textit{code classification} \cite{mou2016convolutional, barchi2021exploration}, \textit{code clone detection} \cite{jiang2007deckard, white2016deep, shobha2021code}, \textit{code summarization} \cite{allamanis2016convolutional, liu2021retrieval}, \textit{method name prediction} \cite{allamanis2015suggesting}, and \textit{source code retrieval} \cite{gu2018deep, ling2021deep}. Representing source code is central to all of these software engineering tasks and is influential in determining the performance of the approaches \cite{zhang2019novel, allamanis2018survey}. 

Many existing works utilize tree and graph-based representations such as \textit{Abstract Syntax Tree} (AST) \cite{white2016deep, jiang2007deckard, zhang2019novel, wang2020learning, wang2020modular, bui2021infercode, li2021fault, kim2021code}, \textit{Control Flow Graph} (CFG) \cite{sun2014detecting, phan2017convolutional}, or \textit{Program Dependency Graph} (PDG) \cite{li2019improving} to represent source code for various tasks. Since programming languages have strict and well-defined grammar, researchers have traditionally used formal methods to process and reason about source code. These formal approaches involve using algorithmic or mathematical techniques over the AST, CFG, or PDG \cite{baxter1998clone, jiang2007deckard, yamaguchi2014modeling}. Due to advancements in deep learning, researchers have started focusing on automatically learning program properties to make software more understandable and maintainable \cite{allamanis2018survey}. These learning-based techniques improved performance over traditional methods \cite{allamanis2018survey}.

Though the learning-based techniques provide superior performance to traditional methods, most of these works are limited to specific aspects of source code. For example, as Table \ref{table:relwork} highlights, AST dominates most of the research, and only a few works use representations such as CFG or PDG. Two critical factors for a learning-based technique are the learning model and the data used to train the model. Data in the current context refers to the code representations used to train the model. We experimented with both these factors to arrive at an optimal model and input data to maximize performance. However, software engineering literature traditionally focused on optimizing the model rather than enhancing the input data to the model. One way to enhance the input data is by using a combination of representations, including various syntactic and semantic code representations. We can include multiple representations in the dataset because, unlike traditional methods, learning-based methods can selectively give importance to those program features that improve the model's performance and neglect the ones that do not. With learning-based approaches, we can offload the responsibility of choosing a suitable representation for a task to the model.

\begin{table}[ht]
\caption{An overview of the literature that utilizes AST / CFG / PDG}
\label{table:relwork}
\centering
\begin{tabular}{C{0.15\linewidth} L{0.2\linewidth} C{0.4\linewidth} L{0.15\linewidth}}
\toprule\toprule
Representation & Work      & Task(s) & Venue \\
\midrule
\multirow{9}{*}{AST} & Deckard \cite{jiang2007deckard} & Code clone detection & ICSE \\
& White et al. \cite{white2016deep}  & Code clone detection & ASE \\
& Wei et al. \cite{wei2017supervised} & Code clone detection & IJCAI \\ 
& TBCNN \cite{mou2016convolutional} & Code classification & AAAI \\ 
& ASTNN \cite{zhang2019novel} & Code classification, Code clone detection & ICSE \\
& MTN \cite{wang2020modular} & Code classification, Code clone detection & TOSEM \\ 
& code2vec \cite{alon2019code2vec} & Method naming & POPL \\ 
& code2seq \cite{alon2018code2seq} & Code summarization, Code captioning & ICLR \\ 
& InferCode \cite{bui2021infercode} & Code clone detection, Code classification, Code clustering, Code search,  Method naming & ICSE \\ [0.8ex]
\midrule
\multirow{2}{*}{CFG} & Sun et al. \cite{sun2014detecting} & Code clone detection & IFIP \\
& Phan et al. \cite{phan2017convolutional} & Software defect prediction & ICTAI \\  [0.8ex]
\midrule
\multirow{1}{*}{PDG} & Li et al. \cite{li2019improving} & Bug detection & OOPSLA \\ [0.8ex]
\midrule
\multirow{3}{=}{Combination of representations} & Allamanis et al. \cite{allamanis2017learning} & Variable name prediction & ICLR \\ & Yamaguchi et al. \cite{yamaguchi2014modeling} & Software vulnerability detection & SSP \\ & Long et al. \cite{long2022multi} & Algorithm classifcation & AAAI \\ [0.8ex]
\bottomrule
\end{tabular}
\end{table}

Few approaches have used different graph structures together to represent source code \cite{yamaguchi2014modeling, allamanis2017learning}. The Code Property Graph (CPG) by Yamaguchi et al. \cite{yamaguchi2014modeling} is one such approach where they integrated CFG, PDG, and AST into a joint \textit{static} data structure to model and detect \textit{software vulnerabilities}. The CPG helps express common code vulnerabilities for specific applications, which can be queried by graph traversal queries \cite{perl2015vccfinder}. Similarly, Allamanis et al. \cite{allamanis2017learning} combine AST, control flow, and data flow edges into a single graph, where these graph edges are particularly designed for the variable name prediction task \cite{alon2019code2vec}. This approach needs execution trace information to create graphs, which is an additional overhead \cite{wang2020modular}. Although these works use semantic graphs and perform well on the intended tasks, they \textit{do not} provide any insight into the effects of using semantic graphs along with AST. Besides, manually choosing the appropriate edges for a task can be laborious and cannot be easily extended for other tasks. 

Thus, in this paper, we try to answer the question: \textit{``How does utilizing multiple source code representations affect the performance of diverse software engineering tasks?"} We do this by extending an existing AST-based approach to semantic graphs like CFG and PDG, then analyzing the impact of these representations on the model's performance. The technique we use in this work involves representing a code snippet as a set of paths extracted from its AST. We selected this path-based technique since there is no need manually design program features for a specific task or language. Moreover, extending this approach to CFG and PDG would allow us to measure the performance boost provided by these semantic graphs. \textit{Code2vec} \cite{alon2019code2vec} is a popular work that uses a similar technique wherein these AST paths are used to train an attention-based neural network. This neural network encodes the code's structural information from individual AST paths by generating method-level code embeddings, which can then be used for various downstream tasks. However, like most existing approaches, their works do not emphasize semantic graphs such as CFG and PDG.

Our work introduces a way to extract paths from semantic graphs CFG and PDG. We utilize paths from both syntactic and semantic representations to work on software engineering tasks. We extend the \textit{code2vec's} attention mechanism to work with paths extracted from multiple code representations, mainly AST, CFG, and PDG. Utilizing the attention mechanism to handle CFG and PDG paths may allow the model to capture the code semantics better, which may not be possible with AST paths alone. This \textit{Aggregate Attention model} gives optimal weights to individual paths extracted from AST to compute a weighted average. It also parallelly calculates the weighted averages of CFG and PDG paths. These three individual weighted averages are concatenated to generate a code embedding. The attention mechanism allows the model to learn how much importance/weight it should give to each path. This mechanism enables the neural network to capture subtle similarities and differences between code snippets, which would be impossible if all paths were given equal weights without using attention.

The reason we take an existing AST-only approach and make it workable with multiple representations is to understand the impact that utilizing multiple representations has on the model's performance. Our goal is \textit{not} to introduce a better learning model but to investigate the possibility of improving the model's performance by experimenting with multiple representations. We chose \textit{code2vec} because it is neither task nor language-dependent and is easily extensible to various tasks like program classification and clone detection. Moreover, it is still widely used, and many recent works are built upon it \cite{alon2018code2seq, compton2020embedding, shi2020pathpair2vec, shi2021toward}. 

\begin{tcolorbox}[width=\linewidth, boxrule=0.5pt]
The core reasoning of this paper is to explore the idea of integrating semantic structures (such as CFG and PDG) with syntactic structures (AST) and study the effect on the performance of various software engineering tasks.
\end{tcolorbox}

As a first step, we evaluated our approach on the task of Method Naming using a custom \textit{C} language dataset of around 730K methods \cite{vagavolu2021mocktail}. The comparative evaluation of \textit{code2vec} and our model shows an increase of \textbf{11\%} in the F1 score on the whole dataset. However, when evaluated on different projects in our dataset, we see an increase of F1 score up to 100\%. This paper is an extension of our preliminary \textit{New Ideas and Emerging Results} (NIER) paper \cite{vagavolu2021mocktail}, which briefly demonstrates this approach on the method naming task. This paper extends our approach to program classification and code clone detection tasks to find the generality across multiple tasks. We also have reimplemented our Path Extractor tool (refer to Section \ref{ssec:pathex}) to extend its functionality for these new tasks.

Our experiments show that our approach improves performance over the AST-only approach for these tasks. Our approach improves the accuracy by \textbf{15.7\%} for program classification and the F1 score by \textbf{9.3\%} for code clone detection. Moreover, our results show that diverse approaches built upon \textit{code2vec} for various tasks \cite{alon2018code2seq, compton2020embedding, shi2020pathpair2vec, shi2021toward} can be enhanced by including paths from CFG and PDG. This paper discusses the additional timing overheads incurred and other implications of our approach. We also discuss threats to the validity of our results and some ways to overcome the shortcomings of our approach. Thus, we believe considering a combination of the syntactic and semantic structures can lead to a new direction in representing source code while also improving existing works that rely solely on AST, CFG, or PDG. We make all our code and data available on GitHub.\footnote{https://github.com/NobleMathews/mocktail-blue-lagoon}

The main contributions of our work are as follows:
\begin{itemize}[itemsep=0.1em]
    \item We introduce a novel way of extracting and utilizing semantic paths from CFG and PDG to represent code.
    \item We employ an attention neural network to learn code embeddings from syntactic and semantic paths. We apply our approach to three tasks: \textit{Method Naming}, \textit{Program Classification}, and \textit{Code Clone Detection}.
    \item We demonstrate that integrating semantic structures (such as CFG and PDG) with syntactic structures (AST) can predominantly improve the performance of these tasks by 11\% (F1), 15.7\% (Accuracy), and 9.3\% (F1), respectively.
\end{itemize}

The remainder of this paper is organized as follows: Section \ref{sec:background} establishes the necessary background. Section \ref{sec:approach} defines semantic paths and explains our attention pipeline. Section \ref{sec:usecase} describes three use cases, demonstrates our experiments, and reports the results. Section \ref{sec:discussion} breaks down the results and discusses the main research questions. Section \ref{sec:threats} discusses the limitations and threats to the validity of our work. Section \ref{sec:relwork} presents the related work. Finally, we conclude the paper in Section \ref{sec:conclusion} with potential implications for further work.

\section{Background}
\label{sec:background}
\subsection{Intermediate Code Representations}
Researchers in program analysis and compiler design have developed multiple intermediate code representations over the years \cite{ferrante1987program, yamaguchi2014modeling}. These representations can also be used in other related fields as they represent the program properties well \cite{yamaguchi2014modeling}. In this work, we use three of the well-known intermediate representations, namely - \textit{Abstract Syntax Tree (AST)}, \textit{Control Flow Graphs (CFG)}, and \textit{Program Dependence Graphs (PDG)}. We chose CFG and PDG to combine with AST because they are the most commonly used representations in literature after AST. We did not select the representations based on their suitability for tasks since each task may require a different semantic representation to achieve the best performance, which would lead to using a different set of representations for each task. But, one of our primary considerations was not to hand-pick the program features and instead let the model decide based on the input code snippets. Moreover, using semantics from multiple dimensions, such as control flow, data dependency, and control dependencies, can still give good results \cite{siow2022learning}. We now define each of these representations below:

\begin{definition}[Abstract Syntax Tree]
It is an ordered tree of source code tokens where the non-terminal nodes represent the statements or operators, and the terminal nodes represent the operands or variables. AST represents the syntax of a program. It is abstract because it does not represent a programming language's actual syntax, instead represents its structure.
\end{definition}

\begin{definition}[Control Flow Graph]
It is a directed graph where the nodes represent predicates and statements, and the edges connecting them indicate the transfer of control between those nodes. It describes the order in which program statements execute. Each edge also has a label that indicates the necessary conditions to execute that path.
\end{definition}

\begin{definition}[Program Dependence Graph]
It is a directed graph where the nodes represent predicates and statements, and the edges connecting them indicate control dependencies and data dependencies between those nodes. PDG's edges are of two types: control dependency edges representing how the result of a predicate influences the variable and data dependency edges that indicate how one variable influences another. Ferrante et al. \cite{ferrante1987program} first introduced this representation for compiler optimization.
\end{definition}

\renewcommand{\thefigure}{1}
\begin{figure}[ht]
    \lstset{language=C}
    \begin{lstlisting}[frame=single]
    void random_function()
    {
        while(!flag)
            if(isValid(x))
                flag = true;
    }
    \end{lstlisting}
    \caption{A sample C function / method}
    \label{fig:code}
\end{figure}

\renewcommand{\thefigure}{2}
\begin{figure}[ht]
    \centering
	\includegraphics[width=\linewidth,height=\textheight, keepaspectratio]{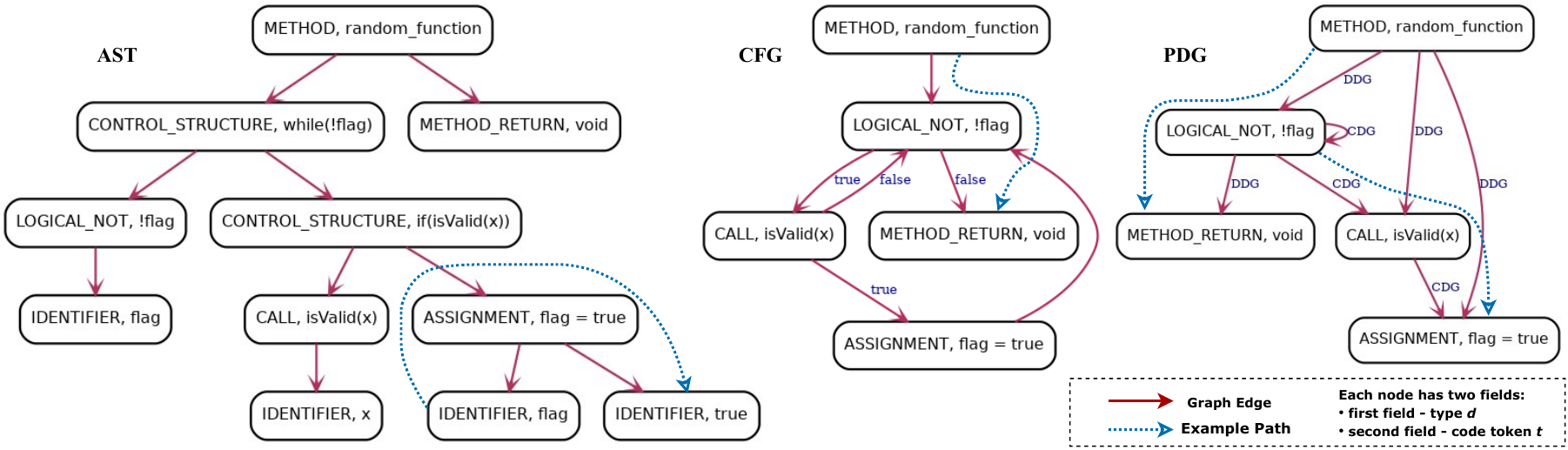}
	\caption{Extracting AST, CFG, and PDG paths from the code snippet in Fig. \ref{fig:code}}
	\label{fig:paths}
\end{figure}

Figure \ref{fig:code} shows a sample C language function called \textit{random\_function}, and Figure \ref{fig:paths} shows the AST, CFG, and PDG of the code snippet in Fig. \ref{fig:code}. Each node in the figure has two fields separated by a comma (,). The first field represents the \textit{type} of the node, the second field represents the actual \textit{code token} associated with the node.

\subsection{Path-based Code Representation}
In their work, Alon et al. \cite{alon2018general} proposed representing a code snippet as a set of paths in its AST. They used these AST paths with Conditional Random Fields (CRF) and evaluated the approach on method naming. Additionally, they introduced the concept of \textit{Path Contexts} in their work. Later, \textit{code2vec} \cite{alon2019code2vec} used these AST paths with an attention-based neural network and gained better results than the CRF approach. We briefly explain the concepts of AST path and path contexts as defined by Alon et al. \cite{alon2018general}:

\begin{definition}[AST Path]
Let $n$ represent a node in an AST, and it has two attributes - the type of node $d$ and the code token $t$. A path between nodes in the AST that begins at a terminal node $n_1$, goes through a series of intermediate non-terminal nodes $n_2$, \ldots, $n_k$, and ends at another terminal $n_{k+1}$ is called an AST path of length $k$. The path $p$ is represented by a sequence of the form: $d_1a_1d_2a_2$\ldots $d_{k}a_{k}d_{k+1}$, where $d_1$, $d_2$, \ldots, $d_{k+1}$ are the types of the nodes $n_1$, $n_2$, \ldots, $n_{k+1}$ respectively, and $a_1$, $a_2$,\ldots, $a_k$ are the directions of movements between path nodes in the AST. Here, $a_i$ $\in$ \{ $\uparrow$, $\downarrow$ \}.
\end{definition}

\begin{definition}[AST Path Context]
It is a tuple $\langle$ $t_1$, $p$, $t_{k+1}$ $\rangle$, where $t_1$ and $t_{k+1}$ are the tokens associated with the AST nodes $n_1$ and $n_{k+1}$ respectively, i.e., terminal nodes of path $p$.
\end{definition}

The intuition behind this approach is that each AST path captures a unique structural template for a set of statements. Consider the code snippet in Fig. \ref{fig:code}. We show the AST for this snippet (with red edges) and an example AST path in Fig. \ref{fig:paths}. This sample AST path represents the statement \textit{flag = true}. Further, any other code snippet with a similar assignment statement in its body (say \textit{a = b}) has a subtree in its AST that is identical to the subtree that represents the statement \textit{flag = true}, regardless of the identifiers used. Hence, both assignment statements can be represented by the same AST path. By this rationale, a set of AST paths can identify the structure of a code snippet. Further, the tokens ($t_1$ and $t_{k+1}$) associated with the two terminal nodes of a path are called the \textit{context words} since they help differentiate the same path that occurs in two different contexts.

\subsection{Attention Mechanism}
\label{ssec:attention}
\textit{Attention} in deep learning is a technique that puts more emphasis/attention on the inputs that help the model make better predictions and less emphasis on the inputs that do not \cite{alon2019code2vec}. An attention layer does this by assigning numerical weights to its inputs to calculate an aggregation (weighted average) of all the inputs.  Attention models are heavily used in NLP tasks like Neural Machine Translation \cite{luong2015effective} and Speech Recognition \cite{bahdanau2016end}. The main advantage of the attention models is that for two similar data instances (but not the same), they generate aggregations that are close in the vector space. The subtle differences in similar inputs are captured through the difference in weights assigned to them. Consider $x_1,$ \ldots$, \, x_n$ to be the inputs to the attention layer. The goal of the attention layer is to learn an attention vector $a$, which can be used to compute the weight $\alpha_i$ for a given input $x_i$. Given the attention vector, weights can be calculated as the normalized dot product of inputs and the attention vector:
\begin{equation}
    \alpha_i = \frac{\exp(a^T \cdot x_i)}{\sum_{j=1}^{n} \exp(a^T \cdot x_j)}
    \label{eq:attention}
\end{equation}

\begin{equation}
    \widetilde{v} = \sum_{i=1}^{n} \alpha_i \cdot x_i
    \label{eq:lincom}
\end{equation}
A weighted average vector $\widetilde{v}$ can then be calculated to generate a single representation for the data instance (Eq \ref{eq:lincom}). The attention vector $a$ will be learned over time to generate better representations for all the instances in the dataset. This type of attention where the weights are calculated using dot product is called \textit{Luong attention} \cite{luong2015effective}.

\section{Approach}
\label{sec:approach}
\subsection{Defining Semantic Path Contexts}
Semantic features such as control flows and data dependencies have been used in some works to solve Software Engineering problems \cite{allamanis2017learning, tufano2018deep}. However, using paths from semantic representations is yet to be explored. AST paths do not capture semantic aspects such as control flow and program dependencies which may be required to achieve optimal performance. To overcome this issue, in this paper, we extend the concept of paths to CFG and PDG:

\begin{definition}[CFG Path]
Let $n$ represent a node in a CFG, and it has two attributes - the type of node $d$ and the code token $t$. A path between nodes in the CFG that begins at the START node $n_1$, goes through a series of intermediate nodes $n_2$, \ldots , $n_k$, and ends at a node $n_{k+1}$ is called a CFG path of length $k$. The last node $n_{k+1}$ can be of two types:
\begin{enumerate}
    \item[] - The END node,
    \item[] - A previously visited intermediate node that represents a loop control structure (i.e. $n_{k+1}$ $\in$ $n_2$, \ldots, $n_k$.)
\end{enumerate}
The path $p$ is represented by a sequence of the form: $d_1$ $\downarrow$ $d_2$ $\downarrow$ \ldots $d_{k}a_{k}d_{k+1}$, where $d_1$, $d_2$, \ldots, $d_{k+1}$ are the types of the nodes $n_1$, $n_2$, \ldots, $n_{k+1}$, and the direction $a_k$ depends upon the last node $n_{k+1}$. Here, $a_k$ is $\downarrow$, if $n_{k+1}$ is END node, $\uparrow$ otherwise.
\end{definition}

\begin{definition}[CFG Path Context]
It is a tuple $\langle$ $t_1$, $p$, $t_{k+1}$ $\rangle$, where $t_1$ and $t_{k+1}$ are the tokens associated with the CFG nodes $n_1$ and $n_{k+1}$ respectively, i.e., terminal nodes of path $p$.
\end{definition}

The START and END nodes are special nodes that represent the start and end of program execution. Each of the CFG paths represents a possible control flow pattern during program execution. To represent loop structures in a CFG, we extract three different paths from it: \begin{itemize}
    \item [1.] A path that ignores the loop and proceeds to the next node (i.e. $n_{k+1}$ = END), 
    \item [2.] A path that goes through the loop only once and proceeds to the next node (i.e. $n_{k+1}$ = END), 
    \item [3.] A path that goes through the loop only once and ends at the loop's start node (i.e. $n_{k+1}$ $\in$ $n_2$, \ldots, $n_k$.)
\end{itemize}

These three paths together represent possible executions of a loop. Hence, for this reason, we have two types of last node $n_{k+1}$ in our definition of the CFG path. Any other constructs like conditionals do not cause loops in CFG and hence are straightforward to handle.

\begin{definition}[PDG Path]
Let $n$ represent a node in a PDG, and it has two attributes - the type of node $d$ and the code token $t$. An edge $e$ in a PDG has a label $l$ associated with it. A PDG path $p$ is a sequence of nodes $n_1$, $n_2$, \ldots, $n_{k+1}$ where all of the edges along the path have the same label $l_p$. The path p is represented by a sequence of the form: $d_1$$a_1$$d_2$$a_2$\ldots $d_{k}a_{k}d_{k+1}$, where $d_1$, $d_2$, \ldots, $d_{k+1}$ are the types of the nodes $n_1$, $n_2$, \ldots, $n_{k+1}$, and $a_1$, $a_2$, \ldots, $a_k$ are the directions of movements between path nodes in the PDG. Here, $a_i$ $\in$ \{ $\uparrow$, $\downarrow$ \} and $l_p$ $\in$ \{ CDG, DDG \}.
\end{definition}

\begin{definition}[PDG Path Context]
It is a tuple $\langle$ $t_1$, $p$, $t_{k+1}$ $\rangle$, where $t_1$ and $t_{k+1}$ are the tokens associated with the PDG nodes $n_1$ and $n_{k+1}$ respectively, i.e., terminal nodes of path $p$.
\end{definition}

Since PDG is a combination of Control Dependency Graph (CDG) and 
Data Dependency Graph (DDG), each PDG path can either represent a control dependence of statements or a data dependence. The labels of edges in a path decide the type of PDG path. Since PDG is a graph (and not a tree), moving up and down does not make sense in PDG. However, we chose to keep all the directions of movements as $\downarrow$. We do this to have a consistent model input format. The concept of CFG and PDG path contexts are similar to AST path contexts. Fig. \ref{fig:paths} depicts the CFG and PDG for the code in Fig. \ref{fig:code} and an example path for each representation.

\subsection{Extracting Path Contexts}
\label{ssec:pathex}
We have developed and used a python-based tool to extract different path contexts from C programs. We first parse each code snippet in the dataset using a platform called Joern \cite{yamaguchi2014modeling} to generate AST, CFG, and PDG. These graphs are exported as .dot files. We then extract paths from these graphs as per the definitions provided earlier. If a snippet has multiple methods, Joern generates all the graphs (AST/CFG/PDG) for each method separately. In such cases, we extract paths from all such graphs and use them to represent the complete code snippet. To account for the high variation in lengths of AST paths, we follow \textit{code2vec's} policy and extract only those with a maximum length of 8 and width of 2. An AST path's width refers to the difference in leaf node indices when all the leaf nodes are indexed sequentially. Further, to avoid high variation in the number of path contexts across code snippets, we limit the number of AST, CFG, and PDG path contexts extracted from a snippet to 200, 10, and 100, respectively. If a graph has more paths than the maximum limit, we randomly sample the maximum number of paths allowed for that representation. Some of the files may have hundreds of paths, and others may have as low as 1, and this variation could create very sparse matrices while training the neural network. So we use these settings to avoid sparse matrices as they could adversely affect the performance. We have set the path limits for AST (200), CFG (10) and PDG (100) based on their average path counts in the dataset (refer Tables \ref{table:mndataset} and \ref{table:ojdataset}).

\renewcommand{\thefigure}{3}
\begin{figure*}
    \centering
    \captionsetup{justification=centering,margin=2cm}
    \includegraphics[width=\textwidth]{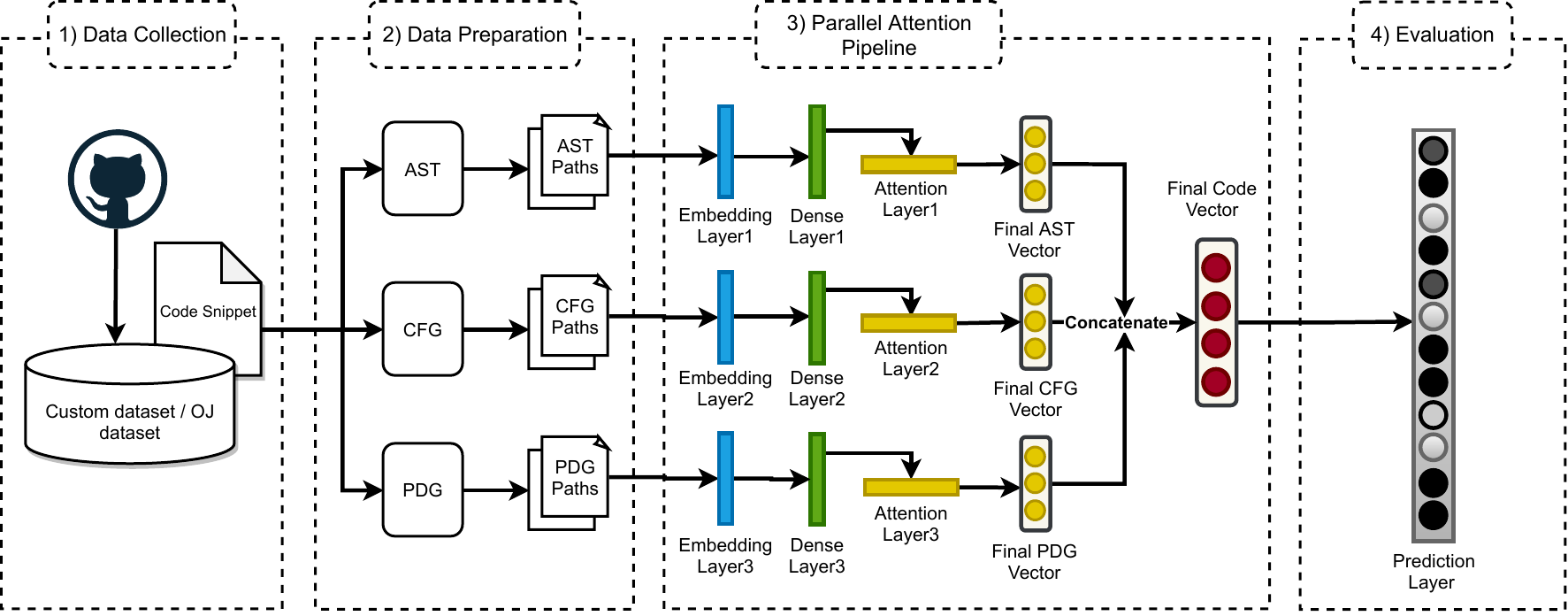}
    \caption{Our pipeline to utilize a combination of Source Code Representations. The specific settings for the final prediction layer are chosen based on the task at hand.}
    \label{fig:model}
\end{figure*}

\subsection{Parallel Attention Pipeline}
Fig. \ref{fig:model} depicts different phases in our approach, and specifically, the third phase shows our parallel attention pipeline. To combine AST, CFG, and PDG paths, we extend the \textit{code2vec} model \cite{alon2019code2vec}, which takes only AST path contexts as input. Our model takes multiple types of path contexts as inputs in a parallel pipeline and finally generates a code vector. We use multiple \textit{code2vec}'s parallelly instead of a single \textit{code2vec} because AST, CFG, and PDG represent different aspects of a program and need to be learned separately, which would allow us to combine their learned vectors later. Consider a code snippet $C$ presented to the model as a bag of path contexts extracted from it.
\begin{equation}
	C = [\{a_1, \ldots, a_{n_1}\}, \{c_1, \ldots, c_{n_2}\} \{p_1, \ldots, p_{n_3}\}],
\end{equation}

\noindent Here $a_i$, $c_j$, and $p_k$ are AST, CFG, and PDG path contexts respectively. Each of the AST path contexts is processed by the model as follows: The three components of a path context (two context words and a path) are passed through embedding layers to generate token and path embeddings of size $D$. These three embeddings are then concatenated and passed through a fully connected (Dense) layer to generate a \textit{context vector} $x$ of size $D$. The \textit{tanh} activation function is used for this dense layer. The primary purpose of this dense layer is to learn to compress the concatenated vector (size $3D$) to generate a context vector (size $D$). This layer learns to perform this compression in a way that gives more importance to a path when it appears with some tokens (context words) and less importance when it appears with others. This helps in differentiating the same path that appears in two different contexts. As a result of this dense layer, context vectors $x_1$, $x_2$, \ldots, $x_{n_1}$ are generated for the AST path-contexts $a_1$, $a_2$, \ldots, $a_{n_1}$ respectively. These context vectors are passed through an attention layer to compute a weighted average, as explained in Section \ref{ssec:attention}. This process is done parallelly for path contexts of each type (AST, CFG, and PDG), and three weighted average vectors $\widetilde{a}$, $\widetilde{c}$, and $\widetilde{p}$ are generated. Then, average vectors are concatenated to generate the final code vector $v$. The code vector $v$ is used to make predictions regarding the code snippet $C$ using another prediction layer. The model is trained to minimize the prediction error. All the context, weighted average, and code vectors are trained and learned concurrently. The prediction layer and the error function are decided based on the task, and we mention the specific details about our experiments in Section \ref{sec:usecase}. We have used Tensorflow's Keras API to build our parallel attention pipeline.

\section{Use Cases}
\label{sec:usecase}
\subsection{Method Naming}
The main goal of method naming is to predict/suggest an accurate name, given the method body. The predicted name should accurately represent the semantics of the method. This is an important problem because having good method names makes the code more readable and maintainable, but poorly named methods can adversely affect the programmers' productivity \cite{allamanis2015suggesting}. Consider $v$ the code vector generated by our model for the input code snippet $C$ and $y_1$, $y_2$, \ldots, $y_N$ are distinct method names found in the training dataset. Our aim now is to use $v$ to predict which of the labels $y_1$, $y_2$, \ldots, $y_N$ is the actual name for the method $C$. After our model produces $v$, we use a fully connected layer with softmax activation function to generate the prediction probability vector $\hat{p}$. Consider $W$ $\in$ $\mathbb{R}^{N \times d}$ as the weight matrix associated with this layer whose rows correspond to the labels $y_1$, $y_2$, \ldots, $y_N$. The probability vector $\hat{p}$ is calculated as follows: 
\begin{equation}
    \hat{p_i} = \frac{\exp(W_i \cdot v)}{\sum_{j=1}^{N} \exp(W_j \cdot v)}
\end{equation}
where $\hat{p_i}$ is the probability of $y_i$ being the method name, and $W_i$ is the row $i$ of $W$. We use the standard cross-entropy loss function for the training: 
    \begin{equation}
        \label{eq:crossloss}
    	loss(t, \hat{p}) = - \sum_{i=1}^{N} t_i \log{ \left( \hat{p_i} \right) },
    \end{equation}
where $t$ is an $N$-dimensional one-hot encoded true label. Then the predicted method name $\hat{y}$ will be the one with highest probability:
    \begin{equation}
        \label{eq:prediction}
        \hat{y} = \argmax_{i} \: (\hat{p_i})
    \end{equation}

\subsubsection{Dataset Preparation}
We opted for a C language dataset to assess our model's performance in the Method Naming task. Our intention is to initially investigate the influence of semantic representations on downstream tasks with an imperative language like C. Once this initial experiments are demonstrated, the work could be further extended to object-oriented languages, such as Java. We could not find any existing C language datasets for method naming, so we decided to collect our own.

As the first step of our dataset collection, we fetched a list of open-source C projects from GitHub and ranked them by popularity. Popularity was quantitatively assessed using a combination of stars and forks, following a standard z-score approach \cite{allamanis2016convolutional, alon2019code2vec}. Repositories containing less than 10,000 methods were excluded from our analysis, as higher method count generally indicates a project with a reasonably rich development history and complexity \cite{omari2020enabling}. Including projects with more than 10K methods also allows us to measure how well the model can generalize when trained on a single repository. Conversely, repositories with over 300K methods were omitted to maintain tractability and prevent any single project from unduly skewing the dataset, thereby ensuring a balanced and manageable scope for analysis. Further, due to the prevalence of similar domains (e.g., operating systems, compilers, databases, and firmware, etc.) in our repository list, we excluded projects of a similar nature at random to retain a more diverse set. To ensure a valid and insightful comparison with \textit{code2vec}, we construct a dataset of comparable size, consisting of roughly 700,000 methods. Ultimately, our curated dataset encompasses 16 open-source C projects, collectively containing 729,218 methods. A breakdown of the repositories considered, including their specific domain and the count of methods, is available in our GitHub repository\footnote{https://github.com/NobleMathews/mocktail-blue-lagoon}.

We preprocess the dataset by dividing all the C source files into individual methods. We then replace any occurrences of the method's name in its body with a special token and store the method name separately. We do this to remove any additional help the model might get from path contexts that contain the method name as a context word. We then normalize all method names by converting them to lowercase and splitting them into subtokens. A special character like \textquotesingle \text{\textbar}\textquotesingle \, separates the subtokens. The normalized name is the true label for the method naming task. Then we extract paths from all the methods as explained in Section \ref{ssec:pathex}. We also filter out invalid methods that do not have at least one path from each representation. In this case, we extract paths at the \textit{method-level} since the task deals with each method individually. These sets of paths are used to train the model to generate \textit{method-level embeddings}. We show the average path count statistics for the dataset in Table \ref{table:mndataset}. In addition to the full dataset, we provide the statistics for 5 out of 16 projects since we use them individually for evaluation.

\begin{table}[ht]
\caption{Average Path Count (per method) for our Dataset}
\label{table:mndataset}
\centering
\begin{tabular}{l  c  c  c  c}
\toprule\toprule
Dataset             & Number of Methods & AST   & CFG & PDG   \\
\midrule
Full C dataset      & 729,218           & 72.8  & 4.4 & 15.8  \\ [0.8ex]
\textit{FFmpeg}     & 15,790            & 131.0 & 5.7 & 32.5  \\ [0.8ex]
\textit{SumatraPDF} & 16,356            & 90.6  & 5.6 & 20.8  \\ [0.8ex]
\textit{KBEngine}   & 21,949            & 78.5  & 5.5 & 20.1  \\ [0.8ex]
\textit{QEMU}       & 39,881            & 92.3  & 4.0 & 17.3  \\ [0.8ex]
\textit{CatBoost}   & 54,365            & 75.1  & 4.6 & 19.7  \\ [0.8ex]
\bottomrule
\end{tabular}
\end{table}

\subsubsection{Training, Evaluation, and Results}
\label{sec:mnresults}
To train and evaluate the model, we shuffle methods from all 16 projects and split them into 649,004 training, 54,691 test, and 25,523 validation methods. Also, we randomly select some projects from the dataset to train and evaluate the model by treating them as individual datasets. We do this to gauge how effectively the model could generalize within individual projects. We train and test the model on each dataset using different combinations of representations (i.e., AST, AST + CFG, AST + CFG + PDG.) We train our model to minimize the cross-entropy loss (Eq. \ref{eq:crossloss}.) We use the Adam optimization algorithm with a batch size of 1024 and a learning rate of 0.001. We use dropout regularization with a dropout value of 0.25 on the context vectors to avoid overfitting. Most of these settings are adopted from \textit{code2vec} to make our model's performance comparison with \textit{code2vec} as fair as possible.

The method naming task is a multi-class classification problem, and for this reason, we use the well-known F1 score as the metric \cite{alon2019code2vec, allamanis2016convolutional}. We briefly describe the Precision, Recall, and F1 score metrics.

\begin{description}
    \item[Precision:] It is the ratio of true positives to all positive predictions. Measures the accuracy of a model's predictions.
    \item[Recall:] It is the ratio of true positives to all actual positives. Measures a model's ability to find all positive instances.
    \item[F1 Score:] The harmonic mean of Precision and Recall. Provides a balanced performance measure for classification.
\end{description}

To measure our model's performance on method naming, we consider the quality of the predicted method name. We calculate the precision and recall over sub-words within the predicted method name. The intuition is that the quality of a predicted method name is primarily determined by the sub-words used to construct it. When a prediction has a high recall, we can infer that model can predict most of the sub-words of the true label (actual method name). When a prediction has high precision, we can say that most of the sub-words in the predicted label are also in the true label.

\begin{table}[ht]
\caption{Results for Method Name Prediction Task}
\label{table:mnresults}
\centering
\begin{tabular}{l  c  c  c  c}
\toprule\toprule
\noalign{\vskip 0.5ex}
\multicolumn{1}{c}{} & \multicolumn{4}{c}{F1 scores} \\ [0.5ex]
\cline{2-5}
\noalign{\vskip 1ex}
Dataset & \parbox[h]{1.3cm}{\centering \textit{code2vec} \\ (AST) } & \parbox[h]{1.8cm}{\centering AST + CFG} & \parbox[h]{1.8cm}{\centering AST + PDG} & \parbox[h]{1.8cm}{\centering AST + CFG + PDG} \\ [0.5ex]
\noalign{\vskip 0.8ex}
\midrule
\noalign{\vskip 0.8ex}
Full C dataset        &  47.5  &  50.5  &  51.3  &  \textbf{52.7} \\ [0.8ex]
\textit{FFmpeg}       &  38.5  &  45.8  &  46.1  &  \textbf{47.3} \\ [0.8ex]
\textit{SumatraPDF}   &  13.7  &  27.8  &  29.4  &  \textbf{33.2} \\ [0.8ex]
\textit{KBEngine}     &  22.9  &  37.3  &  38.9  &  \textbf{41.0} \\ [0.8ex]
\textit{QEMU}         &  26.5  &  34.2  &  34.5  &  \textbf{37.3} \\ [0.8ex]
\textit{CatBoost}     &  37.7  &  46.3  &  47.8  &  \textbf{49.7} \\ [0.8ex]
\bottomrule
\end{tabular}
\end{table}

\begin{quote}\emph{RQ1. How do combinations of representations perform on method naming compared to AST?}\end{quote}
We use \textit{code2vec} as the baseline for all our experiments. Since \textit{code2vec} only uses AST paths, it as a baseline allows us to measure the performance gain achieved using semantic code representations. We did not use any advances after \textit{code2vec} as a baseline since our work is fundamentally an extension of it, and our main goal is to measure the impact of semantic representations but \textbf{not} to design a better learning model (which most of the advances after \textit{code2vec} try to achieve). We summarize our results for the method naming task in Table \ref{table:mnresults}. We can see that by adding additional representations like CFG and PDG, the F1 score increased in all of the cases. For this task, the performance boost given by PDG paths is more than the boost given by the CFG paths. Overall, CFG and PDG increased the F1 score from \textbf{47.5 to 52.7 (11\% increase)} on the full dataset. Furthermore, we observe that the model can capture program properties very well within a project. For example, for the \textit{SumatraPDF} project, the performance boost is more than 100\%. The performance gain for individual projects is more significant than for the entire dataset. Thus, the CFG and PDG paths help the model perform well within a project rather than on the full dataset. This behaviour is expected as the combined dataset has functions from different projects with diverse programming styles and coding patterns. 

Moreover, as Table \ref{table:mndataset} shows, the average AST, CFG, and PDG paths per method in the full dataset is 72.8, 4.4, and 15.8. There are many AST paths per method, even though we limit their number based on the length and width during the path extraction phase. In contrast, though CFG and PDG paths are not limited based on length, the average number of CFG and PDG paths are only 4.4 and 15.8, respectively. This introduces a huge problem of data sparsity on the CFG and PDG pipelines. However, for individual projects, the average number of CFG and PDG paths is much higher, and hence the performance gain is also higher (e.g., more than 100\% increase in F1 for SumatraPDF project compared to 11\% increase for the whole dataset.)

\subsection{Program Classification}
Program classification is a task that is aimed to classify the given code snippet into one of the many classes based on its functionality. It is an important task that is primarily helpful in maintaining huge collections of software \cite{mou2016convolutional, linares2014using}. Consider $v$ the code vector generated by our model for the input code snippet $C$ and $N$ be the total number of classes. The true label $t$ is an N-dimensional one-hot encoded vector. We use a fully connected layer with the softmax activation function as the prediction layer. It takes the code vector $v$ as input to generate the prediction probability vector $\hat{p}$. We use the standard cross-entropy loss function (Eq. \ref{eq:crossloss}) for the training. Then the predicted class label $\hat{y}$ can be calculated as the dimension with the highest value (Eq. \ref{eq:prediction}.)

\subsubsection{Dataset Preparation}
\label{sssec:pcdataprep}
We use the Open Judge (OJ) dataset first introduced by Mou et al. \cite{mou2016convolutional} and has been used in several other works on Program Classification \cite{zhang2019novel, wang2020modular, bui2021infercode}. The dataset is a collection of solutions to 104 different programming problems submitted to an online open judge (OJ). Each problem has 500 different C language solutions, making the dataset a collection of 104 x 500 = 52,000 files divided into 104 classes. We aim to classify the programs that solve the same problem into the same class. 

We prepare the dataset by extracting paths from each source file (Section \ref{ssec:pathex}) and including only those files with at least one path from each representation. Unlike method naming, this task operates with individual files rather than methods. Hence, the set of paths extracted from each program is a \textit{file-level} representation. We can use these file-level representations to train our parallel attention pipeline to generate \textit{file-level embeddings}. We show the average path count statistics for the OJ dataset in Table \ref{table:ojdataset}. One can observe that the average number of paths of each type is significantly higher than in the method naming dataset since each file now can have multiple methods.

\begin{table}[ht]
\caption{Average Path Count (per file) for the OJ Dataset}
\label{table:ojdataset}
\centering
\begin{tabular}{c  c  c  c}
\toprule\toprule
Number of Files & \parbox[h]{1.3cm}{\centering AST}   & \parbox[h]{1.3cm}{\centering CFG} & \parbox[h]{1.3cm}{\centering PDG}   \\
\midrule
52,000           & 175.7  & 29.3 & 74.6  \\ [0.5ex]
\bottomrule
\end{tabular}
\end{table}

\subsubsection{Training, Evaluation, and Results}
\label{sssec:pctrain}
Before training the model, we create Train-Test-Validation splits for the OJ dataset. We split each class (of 500 files) into a 70:20:10 ratio, creating a dataset with 36,400 train samples, 10,400 test samples, and 5,200 validation samples. We train the network using the Adam optimization algorithm with a batch size of 1024 and a learning rate of 0.001, the same model settings and configurations as method naming. In this context, a fully connected layer with the softmax activation function acts as the prediction layer for program classification. We use the classification accuracy on the test dataset as the metric to evaluate the performance as done in other related works \cite{mou2016convolutional, zhang2019novel, wang2020modular}. It is calculated as the percentage of classifications done correctly. 

\begin{quote}\emph{RQ2. How do combinations of representations perform on program classification compared to AST?}\end{quote}
Table \ref{table:pcresults} compares the performance of our approach to the baseline AST. The results follow a similar pattern to that of method naming. We can see that as we include CFG and PDG, the accuracy increases with each addition. The PDG paths significantly contribute to the performance gain than the CFG paths, which may be attributed to the much higher number of PDG paths than the CFG paths (Table \ref{table:ojdataset}). When all three representations are included, the model shows a performance increase of \textbf{15.7\%}.

\begin{table}[ht]
\caption{Results of Program Classification Task}
\label{table:pcresults}
\centering
\begin{tabular}{l c}
\toprule\toprule
\parbox[h]{4cm}{Approach} & \parbox[h]{3.5cm}{\centering Test Accuracy (\%)} \\
\midrule
AST (\textit{code2vec})  & 73.65  \\ [0.8ex]
AST + CFG                & 75.79  \\ [0.8ex]
AST + PDG                & 84.88  \\ [0.8ex]
AST + CFG + PDG          & \textbf{85.23} \\ [0.8ex]
\bottomrule
\end{tabular}
\end{table}

\subsection{Code Clone Detection}
Code clone detection task aims to detect whether a given pair of code snippets are similar. This paper deals with a specific kind of clone pairs called \textit{functional clone pairs} or \textit{\textbf{Type-4} clones} \cite{roy2007survey}. We are not dealing with other types of clone pairs that might be lexically or syntactically similar but lack functional similarity. Type-4 clones typically pose the most intricate detection challenges due to their inherent complexity and are usually the hardest to detect \cite{roy2007survey}. For instance, a pair of code snippets are considered Type-4 clones, irrespective of differences in code tokens and structural composition, as long as they have the same functionality (e.g., the same solution implemented using different algorithms). Owing to our current goal of demonstrating the combination of source code representations for downstream tasks, the detection of Type-1 through Type-3 clones is not within the current scope.

Consider $v_1$ and $v_2$ the code vectors generated by our model for two input code snippets. If the input pair is a true clone pair (i.e., they solve the same problem), its ground truth label ($t$) will be 1, and a false clone pair will have ground truth label -1. Using this convention, we can calculate their functional similarity using the cosine similarity measure:
    \begin{equation}
        \label{eq:cossim}
    	sim(v_1, v_2) = \frac{v_1 \cdot v_2}{|v_1| \cdot |v_2|}
    \end{equation}

The cosine similarity score determines how similar two vectors are in an N-dimensional space, and we later use it to determine a clone pair. To determine if two code snippets constitute a clone pair, we require the vectors of both snippets simultaneously. Consequently, our model cannot be directly used for supervised training aimed at minimizing the error in detecting clone pairs, as it requires modifying the model with additional layers. Instead, we adopted an \textit{unsupervised} code clone detection approach \cite{jiang2007deckard, sajnani2016sourcerercc, bui2021infercode}. It is a commonly used approach where the main idea is to pre-train the model on a different task with the same dataset and use the learned model weights to generate and save the code vectors for all code snippets in the dataset. Then, we form code vector pairs to calculate cosine similarity and determine whether the input code snippets are clones based on whether the similarity score exceeds a threshold (say $\theta$):

\begin{equation}
    \label{eq:threshold}
    \hat{y} = \begin{cases}
                1, & sim(v_1, v_2) > \theta \\
                0, & sim(v_1, v_2) \leqslant \theta
              \end{cases}
\end{equation}

This paper uses program classification as the pretraining task. We chose program classification as the pretraining task because the primary goal in detecting functional clones and classifying programs based on functionality is the same: to capture the functionality of the code snippets.

\subsubsection{Dataset Preparation}
We use the same OJ dataset for the clone detection task as well. The dataset is a collection of solutions to 104 different programming problems submitted to an online open judge (OJ). Each solution is stored as a separate file in the dataset. We prepare the dataset as explained in Section \ref{sssec:pcdataprep} and use it to pretrain the model on the program classification task. However, we use a modified version of the OJ dataset to evaluate the unsupervised code clone detector (Eq \ref{eq:cossim} and \ref{eq:threshold}.) Any two solutions/files from the same category of 500 solutions form a clone pair (i.e., its ground truth label is 1). This is based on the fact that all files belonging a category solve the same problem, and thus, have the same functionality. Conversely, any two files that belong to two different categories solve different problems and \textit{do not} form a clone pair (i.e., its ground truth label is 0). Since we have 104 categories with 500 solutions each, this produces over a trillion possible code pairs, which are not practical to process. So we consider only the first 15 categories with 500 solutions each, which generates 28M code pairs, which is still hard to process. We then randomly sample 50K true clone pairs and 50K false clone pairs to create a final dataset for code clone detection. This modified version of the OJ dataset is called \textit{OJClone}, first introduced by Wei and Li \cite{wei2017supervised} and later adopted by other works \cite{zhang2019novel, bui2021infercode}. Similar to program classification, this task operates at the \textit{file-level}, aiming to evaluate functional similarity between files in the dataset rather than individual functions.

\subsubsection{Training, Evaluation, and Results}
Because we are dealing with unsupervised code clone detection, we first train the model on program classification exactly as in Section \ref{sssec:pctrain} and then save the vectors for all code snippets. Exporting the code vectors makes it easy to create code pairs and to reproduce the results. As explained before, we form 50K clone and 50K non-clone code vector pairs and use cosine similarity to evaluate the model (Eq. \ref{eq:cossim} and \ref{eq:threshold}.) We chose the threshold ($\theta$) as \textbf{0.4} for our clone detector. We chose this value as it allowed our clone detector to achieve optimal performance in all four cases. Since the problem is formulated as a binary classification problem (clone pair or not), we evaluate our approach using Precision, Recall, and F1-score. To calculate these metrics, we treat true clone pairs as positive and false clone pairs as negative samples.

\renewcommand{\thefigure}{4}
\begin{figure*}
    \centering
	\includegraphics[width=\linewidth,height=\textheight, keepaspectratio]{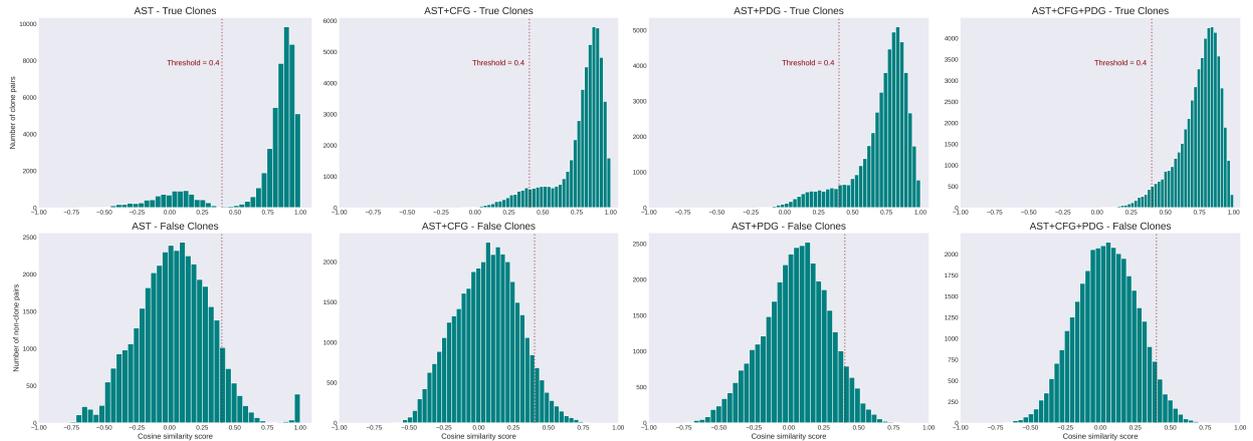}
	\caption{Similarity score distribution for the \textit{OJClone} dataset predicted by four different approaches.}
	\label{fig:ccd}
\end{figure*}

\begin{table}[ht]
\caption{Results of Code Clone Detection Task}
\label{table:ccdresults}
\centering
\begin{tabular}{l c c c}
\toprule\toprule
\parbox[h]{1.6cm}{Approach}& \parbox[h]{1.4cm}{\centering Precision} & \parbox[h]{1.3cm}{\centering Recall} & \parbox[h]{1.3cm}{\centering F1}  \\
\midrule
AST (\textit{code2vec})     &   0.93    &  0.81  & 0.86  \\ [0.8ex]
AST + CFG                   &   0.96    &  0.87  & 0.91  \\ [0.8ex]
AST + PDG                   &   0.95    &  0.86  & 0.90  \\ [0.8ex]
AST + CFG + PDG             &   \textbf{0.96}    &  \textbf{0.92}  & \textbf{0.94}  \\ [0.8ex]
\bottomrule
\end{tabular}
\end{table}

\begin{quote}\emph{RQ3. How do combinations of representations perform on code clone detection compared to AST?}\end{quote}
Fig. \ref{fig:ccd} depicts the cosine similarity distribution for all clone and non-clone pairs. We can see that CFG and PDG cause the similarity distribution of positive samples to shift towards 1, while the similarity distribution of negative samples shifts in the opposite direction. A threshold value of \textbf{0.4} produced the best results in all three cases. We compare the performance of our approach with AST in Table \ref{table:ccdresults}. Once again, the results show that including semantic code representations improves the performance significantly. Though the improvement in precision is only 0.03, this is expected since AST alone achieves an impressive precision of 0.93. However, our approach's main enhancement is the recall, which has increased from 0.81 to 0.92. Overall, including CFG and PDG improves the F1 by \textbf{9\%}. An interesting observation is unlike the previous two tasks, CFG provides a slightly better performance than PDG. Even though the PDG paths are significantly more in number than CFG paths (refer Table \ref{table:ojdataset}), PDG did not provide superior performance gain. This outcome can mean that control flows can capture the behaviour of programs much better than data dependencies. This is understandable since PDG is more inclined towards capturing how a variable depends on another but not how a variable behaves based on other variables.

\section{Discussion}
\label{sec:discussion}
When evaluating how good an approach is to solve a problem, there can be multiple factors to consider, of which we have already discussed an important one: \textit{performance}. Another crucial factor would be the \textit{additional effort} needed compared to existing approaches. As previously discussed, a path-based approach does not manually craft program features but lets the model pick appropriate features for a task. Moreover, the path extraction process can be extended to other languages with minor adjustments, such as replacing the parser and modeling interactions between any language-specific constructs as paths. Since the manual effort is significantly reduced and offloaded to the model, measuring the additional processing overheads incurred in the data preparation and training phases is essential. We analyze the processing overheads in this section and examine our results for further insights. While we acknowledge that there could be other ways of measuring the impact of using multiple source code representations for Software Engineering tasks, we limit the scope of this paper to \textit{performance} (using metrics such as Accuracy, F1, etc.) and \textit{additional overhead incurred} (in terms of computational time overhead).

\begin{quote}\emph{RQ4. What are the additional processing overheads incurred by including multiple representations?}\end{quote}
Table \ref{table:timing} compares the throughputs of four different approaches - AST, AST + CFG, AST + PDG, AST + CFG + PDG. We compare the throughputs at three stages of our pipeline: Path Extraction (Dataset creation), Model Training, and Model Inference. We use a system with two \textit{Intel(R) Xeon(R) CPU E5-2640 v4} chips during the path extraction phase. This system effectively has 40 CPUs, which our path extractor utilizes concurrently to extract paths from multiple source files. We have used a system with three \textit{GeForce GTX 1080 Ti} GPUs for training and evaluating the model. 

Table \ref{table:timing} shows that all three throughputs decrease as we include additional representations. The path extraction throughputs for the method naming task are much higher than for program classification since the former is a method-level task, and thus, samples have significantly fewer paths per representation than the latter. The path extraction throughput decreased by 35.2\% and 34.4\% for both the tasks when CFG and PDG were included. The dataset creation times can be hugely improved by making the path extraction algorithm more efficient and traversing the graphs in-memory (without exporting them to a .dot file).

\begin{table}[ht]
\caption{Comparing the time taken by each approach in different phases of the pipeline}
\label{table:timing}
\centering
\begin{tabular}{lcccccc}
\toprule\toprule
\noalign{\vskip 0.5ex}
\multicolumn{1}{c}{} & \multicolumn{2}{c}{Path Extraction Throughput} & \multicolumn{4}{c}{Model Throughput (samples/second)} \\ [0.5ex]
\cmidrule(l){4-7}
\noalign{\vskip 0.5ex}
\multicolumn{1}{c}{} & \multicolumn{2}{c}{(samples/minute)} & \multicolumn{2}{c}{During Training} & \multicolumn{2}{c}{During Inference} \\ [0.5ex]
\cmidrule(l){2-3}
\cmidrule(l){4-5}
\cmidrule(l){6-7}
\noalign{\vskip 1ex}
\parbox[h]{1.6cm}{\centering Representations \\ used}
& \parbox[h]{1.55cm}{\centering Method Naming} & \parbox[h]{2cm}{\centering Program Classification} & \parbox[h]{1.55cm}{\centering Method Naming} & \parbox[h]{2cm}{\centering Program Classification} & \parbox[h]{1.55cm}{\centering Method Naming} & \parbox[h]{2cm}{\centering Program Classification} \\  [0.5ex]
\noalign{\vskip 0.8ex}
\midrule
\noalign{\vskip 0.8ex}
AST             & 2272 & 29 & 268 & 846 & 858 & 4110 \\  [0.8ex]
AST + CFG       & 1827 & 23 & 179 & 535 & 694 & 3827 \\  [0.8ex]
AST + PDG       & 1538 & 22 & 164 & 461 & 619 & 2854 \\  [0.8ex]
AST + CFG + PDG & 1470 & 19 & 126 & 317 & 563 & 2722 \\  [0.8ex]
\bottomrule
\end{tabular}
\end{table}

The model throughput during the training phase is quite affected when all three representations are used. The training throughput decreased by 53\% and 62.5\% for method naming and program classification tasks. The throughput for program classification is more affected because the average number of CFG and PDG paths at the file level is much higher than at the method level (refer to Tables \ref{table:mndataset} and \ref{table:ojdataset}), and hence it takes more time for the model to process a code snippet for program classification. The program classification task's training throughputs are much greater than the method naming task. One might think this is a contradiction since a file-level task has more paths per sample than a method-level task, but other factors also affect model throughput, for example, model size and dataset size. The dataset for method naming is much larger than program classification, which increases its model size and thus decreases the training throughput.

Though the model's inference throughput was also affected when we used multiple representations, the throughput is still acceptable. To get a much clearer picture, we should look at the total time taken during inference, including time taken for path extraction and model inference time. For method naming, the average total inference time per sample comes out to be \textit{0.027 seconds} using AST and \textit{0.043 seconds} using all three representations (59\% increase). Similarly, average inference times for program classification are \textit{2.07} and \textit{3.16 seconds}, respectively (52.6\% increase). Most of the increase in time comes from the path extraction phase. As mentioned, one can decrease path extraction times using better algorithms, avoiding expensive file operations, etc.

\begin{quote}\emph{RQ5. What are the implications of our results? and how does each representation contribute to the model's performance?}\end{quote}
Our experiments show that each semantic representation, when used with AST, only increases the model's performance but never deteriorates it. This observation indicates that AST does not provide a complete picture, and we might need additional representations to capture program features (semantics) effectively. However, some representations may be much effective than others for a given task and dataset. Consider Table \ref{table:perfcompare}, where we compare the performance gain provided by CFG and PDG for all three tasks. We measure performance gain as the increase in the performance metric when CFG/PDG is included. We can observe that the combination AST + PDG is more effective than AST + CFG for method naming and classification tasks. This may be due to the higher number of paths in PDG than CFG, which leads to PDG providing more features for the model to generalize. Although the combination AST + CFG gives a slightly better performance for clone detection than AST + PDG, the results are close. This variation suggests that selecting the most suitable combination of code representations is largely task-dependent.

\begin{table}[ht]
\caption{Comparison of performance gain provided by CFG and PDG}
\centering
\begin{tabular}{l c c c}
\toprule\toprule
\parbox[h]{1.6cm}{Task}& \parbox[h]{1cm}{\centering AST} & \parbox[h]{1.9cm}{\centering AST + CFG} & \parbox[h]{1.9cm}{\centering AST + PDG}  \\ [0.5ex]
\midrule
Method Naming (F1)        &  47.5   &  50.5   &  \textbf{51.3}   \\ [0.8ex]
Classification (Accuracy) &  73.65  &  75.79  &  \textbf{84.88}  \\ [0.8ex]
Clone Detection (F1)      &  0.86   &  \textbf{0.91}   &  0.90   \\ [0.8ex]
\bottomrule
\end{tabular}
\label{table:perfcompare}
\end{table}

Furthermore, our results indicate that using all three representations may not always yield a significant performance boost compared to a more focused subset. For example, in the case of the program classification task, AST + PDG proves to be an optimal combination, delivering nearly equivalent performance to AST + CFG + PDG, but with higher training and inference throughputs of 461 and 2854 samples/second, compared to 317 and 2722 samples/second for AST+CFG+PDG. This suggests that customizing code representations for specific software engineering tasks has the potential to reduce processing times while maintaining good performance. While we have observed promising results within the context of our study, further research on a broader range of code bases and tasks is essential to validate these observations and establish generalized conclusions.

\section{Threats to Validity}
\label{sec:threats}
\subsection{External Validity}
\textit{Language Generalization:} Our study primarily focuses on the C programming language, and the conclusions drawn may not directly generalize to other languages. Hence, there is a need to adapt our approach for each language, taking into account their distinct characteristics.

\textit{Dataset Bias:} The diversity of the dataset used for training and evaluation can significantly impact the results. While we tried to diversify the dataset with projects from different domains, it still may not fully represent the broader landscape of software projects. Further exploration using larger datasets, broader in scope, and encompassing languages beyond C is crucial to draw generalizable conclusions.

\subsection{Internal Validity}
\textit{Data Sparsity:} Though our approach significantly boosts the performance for all three tasks, the data sparsity problem discussed in Section \ref{sec:mnresults} can potentially impact the model negatively when used on a larger scale. One possible solution to address this sparsity is to devise more unique ways to formulate and extract semantic features from these representations, such as paths that capture interactions between files and modules. We can also use additional semantic features like \textit{data flow} in addition to control flows (CFG) and program dependencies (PDG). One downside of this approach is the increased effort for pre-processing and training, which can be decided as a trade-off for a respective downstream task. We are essentially extracting more information from a program in the form of paths and training the model rather than extracting limited information and requiring more training data to achieve the same performance.

\textit{Parameter Settings:} To ensure a fair comparision with code2vec, we have adopted code2vec's network hyperparameters, including learning rates, batch sizes, and dropout rates. Extensive hyperparameter tuning is essential to achieve optimal results with the aggregate attention model. Additionally, limiting maximum path lengths and widths to 8 and 2 (refer \ref{ssec:pathex}) may potentially exclude relevant information and affect the quality of extracted features. Similarly, using threshold values on the number of paths extracted from a code snippet (200 AST paths, 10 CFG paths, and 100 PDG paths) derived from average path counts within the dataset may not universally accommodate diverse code structures, and can potentially impact representation quality. Furthermore, while avoiding sparse matrices in neural network training, the strict enforcement of path count limits may inadvertently lead to information loss in code snippets with inherently larger path counts. A detailed sensitivity analysis could be performed to optimise and fine-tune parameters such as path lengths, widths and maximum path counts. Such analysis would involve assessing the impact of these parameter choices on the performance of the Aggregate attention model on downstream tasks. However, in the current work we limit ourselves towards exploring the impact of using multiple code representations in software engineering downstream tasks.

\subsection{Construct Validity}
\textit{Automation Trade-Off:} Another drawback to our approach is that since we are not manually designing program features to represent code, some unusual cases like \textit{inline assembly} may not be appropriately handled while extracting paths. Accommodating such unique scenarios through automation is a trade-off that requires continued exploration and refinement.

\section{Related Work}
\label{sec:relwork}
\subsection{Source Code Representation}
Traditionally, researchers have expressed source code as a sequence of tokens to address important software engineering tasks \cite{kamiya2002ccfinder, zhou2012should, sajnani2016sourcerercc}. \textit{SourcererCC} \cite{sajnani2016sourcerercc} creates a partial index for source code by using code tokens and then uses it to detect code clones. For the task of bug localization, Zhou et al. \cite{zhou2012should} treat source code files as a text corpus to find the similarity between each file and the bug report.

Representing structural information of source code using Abstract Syntax Trees (AST) has emerged as a critical approach, capturing both lexical and syntactic information. ASTs have been used extensively in the literature \cite{white2016deep, jiang2007deckard, zhang2019novel, wang2020learning, bui2021infercode, li2021fault, kim2021code}. For example, \textit{Deckard} \cite{jiang2007deckard} introduces an algorithm for identifying similar sub-trees of two ASTs to detect code clones. Researchers have also tried to capture syntactic information using deep learning models like RNN \cite{white2016deep} or Tree-LSTM \cite{wei2017supervised} on ASTs to detect code clones. Mou et al. \cite{mou2016convolutional} use a custom Tree-based Convolutional Neural Network (TBCNN) on ASTs to learn vector representations of code snippets. In their work, Zhang et al. \cite{zhang2019novel} extract the sub-trees from AST and feed them to an AST-based neural network (ASTNN) to generate code vectors that can capture the sequential dependency of code statements. Li et al. \cite{li2019improving} utilize the global \textit{program dependencies} of source code along with the local ASTs to predict bugs. Alon et al. \cite{alon2018general, alon2019code2vec} introduce the AST path-based approach for representing source code and various learning models to generate code vectors using AST paths. \textit{Code2seq} \cite{alon2018code2seq} adopts a similar strategy to \textit{code2vec} for the task of Neural Machine Translation. More recently, Wang et al. \cite{wang2020learning} introduced heterogeneous program graphs by including additional type information for nodes and edges in an AST and used GNNs to learn program properties. In another work, Wang et al. \cite{wang2020modular} use a modular tree-based network to detect the semantic difference in programs based on their ASTs. \textit{InferCode} \cite{bui2021infercode} use the subtrees of an AST as the training labels and train a TBCNN in a \textit{self-supervised} way. This way, the generated code vectors are not tied to a specific task. Xiao et al. \cite{xiao2022path} used a new notion of \textit{path context} and introduced the path context augmented network (PCAN) to learn code vectors.

Researchers also explored the usage of \textit{Data Flow Graphs} to capture source code's structure \cite{guo2020graphcodebert, vytovtov2019unsupervised}. \textit{GraphCodeBERT} \cite{guo2020graphcodebert} introduced a transformer-based model that uses the data flow in programs to learn the representations. Instead of taking the syntactic-level code structure like an AST, these approaches use the data flow to capture the inherent code structure.

Some of the works have combined different graph structures; for example, Allamani et al. \cite{allamanis2017learning} represent the program as a directed graph of code tokens with different labeled edges like syntax tree, control flow, and data flow for predicting variable and method names. The Code Property Graph (CPG) by Yamaguchi et al. \cite{yamaguchi2014modeling} is the most relevant work to our approach where they create a \textit{static} combination of AST, CFG, and PDG for the task of vulnerability detection. They show its effectiveness by finding 18 previously undiscovered vulnerabilities in the Linux kernel's source code. Zhang et al. \cite{zhang2020exploiting} constructed a code knowledge graph and used a bi-attention layer neural network to detect bugs. More recently, Long et al. \cite{long2022multi} introduced a multiview graph using data-flow, control-flow, read-write graphs to obtain multiple perspectives about source code and employed a GGNN to extract information from them.

Many works have also built upon the \textit{code2vec} model for various downstream tasks \cite{compton2020embedding, shi2020pathpair2vec, shi2021toward}. Compton et al. \cite{compton2020embedding} extend the \textit{code2vec} model to Java classes by aggregating different method embeddings found in a class. Shi et al. \cite{shi2020pathpair2vec} use the \textit{code2vec} model on pairs of AST paths as input for the task of defect prediction. They show an increase in performance over the state-of-the-art model by 17\%. In another work, Shi et al. \cite{shi2021toward} use the \textit{code2vec} model for discovering misconceptions in computing assignments. Our results show that the works built upon \textit{code2vec} can be enhanced by including semantic paths.

\subsection{Method Naming}
Allamanis et al. \cite{allamanis2015suggesting} is the first work that explicitly proposes a solution to the method naming problem to the best of our knowledge. They use a log-bilinear neural network to map the method names to a high-dimensional continuous space such that semantically similar names are closer to each other in the space. As the method's name usually indicates its semantics, this problem is a special case of code summarization. With this intuition, some of the works on code summarization use method naming to evaluate the approach. For example, Allamanis et al. \cite{allamanis2016convolutional} introduce a convolutional attention neural network that takes code tokens as input to detect program features in a context-dependent way and produce a method name for that code snippet. Alon et al. \cite{alon2018general} use AST paths as inputs to Conditional Random Fields (CRF) to predict a method name for the input code snippet. They improved the performance in their subsequent work, \textit{code2vec} \cite{alon2019code2vec}, using AST paths as input to an attention-based neural network. Later, \textit{code2seq} \cite{alon2018code2seq} followed a similar path-based approach with a different model and achieved a significant performance boost compared to previous works. Recently, Wang et al. \cite{wang2020learning} used GNNs on their custom program graphs to predict method names.

\subsection{Program Classification}
Several works have tried to classify source code based on the programming language used \cite{van2016software, gilda2017source}, authorship \cite{frantzeskou2008examining}, domain \cite{linares2014using}, or as in our case, its functionality \cite{mou2016convolutional, zhang2019novel, wang2020modular}. Mou et al. \cite{mou2016convolutional} are one of the first works that use a learning model (TBCNN) to classify programs based on functionality. Later, Zhang et al. \cite{zhang2019novel} proposed ASTNN to overcome the problem of vanishing gradients from which previous deep learning techniques suffered. Wang et al. \cite{wang2020modular} proposed a model (MTN) with multiple neural modules to deal with different AST semantic units. Their approach showed an accuracy improvement of 1.8\% over TBCNN. More recently, \textit{InferCode} \cite{bui2021infercode} used \textit{self-supervised} learning to improve the performance of TBCNN by 4\%.

\subsection{Code Clone Detection}
Researchers have been actively studying code clone detection due to its applications in software engineering \cite{roy2007survey}. Many works have suggested a wide range of approaches, from token-based \cite{kamiya2002ccfinder, sajnani2016sourcerercc} or tree-based techniques \cite{baxter1998clone, jiang2007deckard} to supervised \cite{wei2017supervised, saini2018oreo, zhang2019novel, fang2020functional} and unsupervised \cite{white2016deep, bui2021infercode} deep learning methods. Baxter et al. \cite{baxter1998clone} detect code clones by generating ASTs for input code snippets and comparing their subtrees. This approach is one of the first attempts to solve this problem without using string matching. \textit{CCFinder} \cite{kamiya2002ccfinder} creates a regularized token sequence from the programs to detect duplicate code. White et al.'s \cite{white2016deep} deep learning based technique is one of the earlier attempts to automatically learn program features using a Recursive Neural Network on ASTs to detect code clones. Wei and Li \cite{wei2017supervised} propose a deep learning framework that detects clones by learning features using an AST-based LSTM. Several supervised learning techniques like Oreo \cite{saini2018oreo}, ASTNN \cite{zhang2019novel}, MTN \cite{wang2020modular}, and Fang et al.'s \textit{fusion learning} \cite{fang2020functional} have also shown significant improvements over traditional techniques.

\section{Conclusion and Future Work}
\label{sec:conclusion}
Moving away from the predominantly common approach of using AST for downstream tasks, in this work, we explored the idea of integrating AST with semantic code representations (CFG / PDG) to measure the impact on Software Engineering tasks. Towards this goal, we extended an AST path-based technique and adapted it to include CFG and PDG path contexts. This allowed us to measure the impact of using multiple code representations. We evaluated our approach on the method naming task with different sets of data, first with the full C dataset of 16 projects and then with some individual projects. By including CFG and PDG path contexts, we demonstrate that the model outperforms \textit{code2vec} by 11\% on the full dataset and up to 100\% on individual projects. The performance boost observed for individual projects is much more significant than for the entire dataset, potentially owing to the higher variation in the number of AST, CFG, and PDG paths. To test whether this approach has the potential to be generalized for multiple software engineering tasks, we also evaluated the approach on program classification and code clone detection. The combination AST+CFG+PDG outperforms \textit{code2vec} in these two tasks by 15.7\% (Accuracy) and 9.3\% (F1), respectively. We also measured the impact of multiple representations by measuring the additional computational overhead incurred.

While our initial findings are promising, we see immense scope for further analysis towards generalization. One interesting direction is to extend the approach to more programming languages. Specifically, the approach could be extended to object-oriented languages such as Java to capture the control flow and dependencies between objects and classes. Analyzing the impact of different language characteristics on the effectiveness of semantic representations and model performance can provide valuable insights. To make our approach more robust and widely applicable, it is important to work with more diverse and extensive datasets. This can potentially provide a better understanding of the impact of semantic representations in different software development scenarios, taking into account various coding styles and practices. Another research direction is to explore other semantic representations such as \textit{Data Flow Graph}.

Though our primary intent with this work is not to propose an efficient learning model, our approach can still be explored for other tasks, such as \textit{bug localization} and \textit{code generation}. The proposed approach itself can be improved by conducting comprehensive hyperparameter optimization and sensitivity analyses to fine-tune model configurations. This can help achieve the optimal performance with our approach. The proposed approach can also be leveraged to improve the performance of several works built upon \textit{code2vec}. Moreover, the way these representations are utilized can be modified to suit the task at hand. Also, the proposed path-based approach itself could be further investigated to find the \textbf{\textit{optimal combinations of code representations}} for various downstream tasks. Another direction of future work in our approach is to solve the data sparsity problem during training, as discussed in Section \ref{sec:discussion}. 

We anticipate that this study can motivate researchers to replicate existing approaches by integrating syntactic and semantic representations. In addition, we see that the proposed approach can lay the groundwork to spin off multiple novel code representations for various sub-domains of Software Engineering while efficiently leveraging advances in AI and Programming Language Processing.


\bibliographystyle{elsarticle-num} 
\bibliography{mocktail}
\end{document}